\documentclass[preprint,aps,showpacs]{revtex4}
\begin{document}
\preprint{UCI-TR 2005-4}
\title{Coordinate noncommutativity as low energy effective dynamics}
\author{Myron Bander\footnote{Electronic address: mbander@uci.edu}
}
\affiliation{
Department of Physics and Astronomy, University of California, Irvine,
California 92697-4575}

\date{January\ \ \ 2005}

\begin{abstract} Coordinate noncommutativity, rather than being
introduced through deformations of operator products, is achieved by
coupling an auxiliary system with large energy excitations to the one
of interest. Integrating out the auxiliary dynamics, or equivalently
taking ground state expectation values, leads to the desired
coordinate noncommutativity. The product responsible for this
noncommutativity is different from the Groenewold-Moyal one. For
products of operators at unequal times, this procedure differs from the
normal, commutative one, for time differences smaller than ones
characterized by the auxiliary system; for larger times the operator
algebra reverts to the usual one.
\end{abstract}
\maketitle
Noncommutativity between space coordinates is most commonly introduced
by way of the Groenewold-Moyal \cite{GM} star product deformation of
the ordinary product; it suffices to define this for exponential
functions as the product of other functions is obtained from Fourier
transforms,
\begin{equation}\label{GMstar}
\exp i{\bf k}\cdot{\bf r}\star \exp i{\bf q}\cdot{\bf r}=
  \exp i\left[({\bf k}+{\bf q})\cdot{\bf r}\right]\exp
  \left(\frac{i}{2}\theta_{ab} 
    k_aq_b\right)\, .
\end{equation}
As all vectors are two dimensional, we may set
$\theta_{ab}=\theta\epsilon_{ab}$, where $\theta$ determines the
magnitude of the noncommutativity. The above is the unique {\em
associative} deformation \cite{unique} of the ordinary product. Such
noncommutative geometries arise on D-branes in magnetic backgrounds
\cite{Seiberg:1999vs}. Noncommutativity induced by magnetic fields can
be applied to the quantum mechanics of point particles
\cite{Bigatti:1999iz,Jackiw:2001dj,Jackiw:2002wd}. In the latter
case it is noted that in the dynamics of a charged particle restricted
to the lowest Landau level of a uniform magnetic field the momenta
decouple and one of the space coordinates, $x$, becomes the canonical
momentum of the other one, $y$; the resultant commutator of $x$ and
$y$ is inversely proportional to the applied field.

An advantage of this second approach is that an ordinary product
algebra is maintained. However, the disappearance of the momenta from
the dynamics of a charged system in a strong magnetic field prevents
us from applying this method to problems where we wish to keep both
momenta and coordinates, with commutation relations $[p_a,p_b]=0$ and
$[p_a,r_b]=-i\delta_{ab}$ ($\hbar$ is set to 1) while the space-space
commutator becomes
\begin{equation}\label{spspcom}
[r_a,r_b]=-i\theta\epsilon_{ab}\ . 
\end{equation}

In this work we achieve this goal by correlating the dynamics of the
system of interest to an auxiliary one and then integrating out the
second system. The resultant effective theory, valid for energies
lower than the excitation energies of the auxiliary system, yields the
desired non trivial space-space commutator. The price paid is that the
effective low energy theory is nonlocal in time. It is, of course,
possible to forgo the auxiliary system and postulate the minimum
effective theory as needed to produce noncommutativity. We begin with
this approach and then motivate it and make it more precise by
introducing the auxiliary system leading to the modified dynamics as a
low energy effective theory. The minimal effective theory reproduces
the Groenewold-Moyal product. The product resulting from the more
fundamental approach is significantly different; the commutators in
the two approaches are identical.

Starting with the action for a problem with normal commuting coordinates,
\begin{equation}\label{S0}
S_0=\int dt \left[{\bf p}\cdot {\dot{\bf r}} -H_0({\bf p},{\bf
r})\right]\, ,
\end{equation}
the commutation relation (\ref{spspcom}) can be implemented by adding
to $S_0$ a term, nonlocal in time,
\begin{equation}\label{Snc1}
S_{\rm nc}=\frac{\epsilon_{ab}}{4}\int dt\,  dt'{\dot p}_a(t)
 \epsilon(t-t')\Theta(|t-t'|){\dot p}_b(t')\, ;
\end{equation} where $\Theta(|t|)$ is a function that goes to zero for
$|t|$ much larger than some characteristic time $\tau=1/\omega$ and
$\Theta(0)=\theta$. $\tau$ is taken to be much smaller than any time
relevant to $H_0$; we assume that we are dealing with momenta $p$ and
potentials $V(x,y)$ that satisfy
\begin{equation}\label{momcond}
\tau\frac{\bf p}{m}\cdot\mbox{{\boldmath$\nabla$}}V(x,y)<<V(x,y)\, ;
\end{equation}
namely for space intervals traversed in times less than $\tau$,
the potential dos not vary much. 

Expectation values of products of position operators are obtained
from the generating functional $Z[{\bf j}(t)]$ given by the path
integral
\begin{equation}\label{genfunc1}
Z[{\bf j}(t)]=\int [d{\bf p}][d{\bf r}]\exp i\left[S_0+S_{\rm nc}+
   \int dt {\bf j}(t)\cdot{\bf r}(t)\right]\, ;
\end{equation}
the time ordered product of the $r_a(t)$'s is
\begin{equation}\label{Tproduct1}
\langle T\left[r_a(t_a)r_b(t_b)\cdots\right]\rangle
  =\frac{1}{Z[0]}\frac{-i\delta}{\delta j_a(t_a)}
    \frac{-i\delta}{\delta j_b(t_b)}\cdots Z[{\bf j}(t)]\, ,
\end{equation}
and the expectation value is taken in the joint ground
of $H+H_{\rm nc}$.  Condition (\ref{momcond}) permits an explicit
evaluation of the modification to (\ref{genfunc1}) due to $S_{\rm nc}$
\begin{equation}\label{genfunc2}
Z[{\bf j}(t)]=\exp i\left[\frac{\epsilon_{ab}}{4}\int dt\,  dt' j_a(t)
 \epsilon(t-t')\Theta(|t-t'|) j_b(t')+{\rm usual\ terms}\right]\, .
\end{equation}
The usual terms do not yield in any coordinate noncommutativity; the
term due to $S_{\rm nc}$ results in
\begin{eqnarray}\label{Tpoduct2}
\ldots x(t+\epsilon)y(t)\ldots&=&\ldots\frac{-i}{2}\Theta
(\epsilon)\ldots
\, ;\nonumber\\
\ldots y(t+\epsilon)x(t)\ldots&=&\ldots\frac{+i}{2}\Theta
(\epsilon)\ldots\ , 
\end{eqnarray}
with the dots indicating other operators at times later and earlier
than $t$. The limit $\epsilon\rightarrow 0$ this yields the desired
commutation relation. (It is interesting to note that in the path
integral formulation the usual momentum-coordinate commutation
relations also arise \cite{Feynman} by considering time ordered
products of $p(t)=m{\dot x}(t)$ and $x(t')$ and letting $t'\rightarrow
t$.) Similarly we find
\begin{equation}\label{Moyal2}
\lim_{\epsilon\rightarrow 0}\, \langle\exp{i{\bf k}\cdot{\bf
r}(t+\epsilon)} 
    \exp{i{\bf q}\cdot{\bf r}(t)}\rangle = \langle\exp{i({\bf k}+{\bf
    q})\cdot 
    {\bf
    r}(t)}\rangle\exp{\frac{i\theta}{2}\epsilon_{ab}k_aq_b}\, ,
\end{equation}
as stated earlier, the usual Groenewold-Moyal product.

Can the action $S_{\rm nc}$ be obtained from dynamics local in time? 
To this end we will couple the system of interest, $H_0$, to a one
dimensional auxiliary one with a momentum operator $\rho_x$ and a
position operator $\rho_y$ with
\begin{equation}\label{rhocomm}
[\rho_x,\rho_y]=-i\, ;
\end{equation}
it is for later convenience that we denote the momentum conjugate to
$\rho_y$ by $\rho_x$. For simplicity the Hamiltonian for this
auxiliary system is taken as
\begin{equation}\label{Hosc}
H_{\rm aux}=\omega(\rho_x^2+\rho_y^2)/2\, ;
\end{equation}
$\omega$ will be chosen larger than
the energies of interest in $H_0$. The coupling is accomplished by
replacing the operators ${\bf r}$ by ${\bf R}=(X,Y)$ with
\begin{eqnarray}\label{rtoR}
X&=&x+\sqrt{\theta}\rho_x\nonumber\\
Y&=&y+\sqrt{\theta}\rho_y\; .
\end{eqnarray}
This coupling is unusual in that while the position operator $y$ 
displaced by a term proportional to $\rho_y$, the other position
operator $x$ is displaced by a term proportional to the conjugate
momentum $\rho_x$. Further on we return to this point. For large
$\omega$ the excitations of $H_{\rm aux}$ will be limited insuring
that $\langle\rho_x^2\rangle$ and $\langle\rho_y^2\rangle$ are of order
one and in $\langle({\bf R}-{\bf r})^2\rangle\sim\theta$.  In addition
to the replacement of ${\bf r}$ by ${\bf R}$ in $H_0$, the source term
also changes to ${\bf j}(t)\cdot {\bf R}(t)$. It is clear that
$[X,Y]=-i\theta$. The analog of (\ref{genfunc1}) is
\begin{equation}\label{genfunc3}
Z[{\bf j}(t)]=\int [d{\bf p}][d{\bf r}][d{\mbox{\boldmath $\rho$}}]
\exp i\int dt \left[{\bf p}\cdot
{\dot{\bf r}} +\rho_x{\dot\rho_y} -H({\bf p},{\bf
R})-H_{\rm aux}({\mbox{\boldmath $\rho$}})+{\bf j}(t)\cdot{\bf
R}(t)\right]\, . 
\end{equation}
After changing the variables of integration from ${\bf r}$ to
${\bf R}$, the integration over $\rho_x$ and $\rho_y$ maybe performed
yielding
\begin{equation}\label{genfunc4}
Z[{\bf j}(t)]=\int [d{\bf p}][d{\bf R}]\exp i\int dt\left[ 
{\bf p}\cdot{\dot{\bf R}}-H({\bf p},{\bf R})+
{\bf j}(t)\cdot{\bf R}(t)+S_{\rm nc}\right]\, ,
\end{equation}
with
\begin{equation}\label{Snc3}
S_{\rm nc}=\frac{\epsilon_{ab}}{4}\theta\int dt\, dt'
{\dot p}_a(t)\epsilon(t-t') e^{-i\omega|t-t'|}{\dot p}_b(t')+
\frac{i\theta}{4}\int dt\, dt' {\dot{\bf p}}(t)\cdot{\dot{\bf p}}(t')
e^{-i\omega|t-t'|}\, .
\end{equation}
The first term above is of the form proposed in (\ref{Snc1}) with
$\Theta(|t|)=\theta\exp(-i\omega t)$. While the second term does not
contributing to coordinate noncommutativity it has the interesting
consequence of making the resulting prescription for the product of
operators different from the star product one. The noncommutative
product (\ref{Moyal2}) becomes
\begin{eqnarray}\label{Moyal3}
\lim_{\epsilon\rightarrow 0} \exp\left[{i{\bf k}\cdot{\bf
r}(t+\epsilon)}\right]
    &\times& \exp\left[{i{\bf q}\cdot{\bf r}(t)}\right]\nonumber\\ 
&=& \exp\left[{i({\bf k}+{\bf q})\cdot
    {\bf r}(t)}\right]\exp\left(\frac{i\theta}{2}\epsilon_{ab}k_aq_b\right)
     \exp\left[{-\frac{\theta}{4}({\bf k}+{\bf q})^2}\right]\, .
\end{eqnarray} The extra term, compared to (\ref{Moyal2}), has its
origin in the second term of (\ref{Snc3}) and may be understood by
looking at an operator formulation of this procedure. Position
dependent operators, $O({\bf r})=\int d{\bf q}\, {\tilde O}(\bf
q)\exp{i{\bf q}\cdot{\bf r}}$, are replaced by (see (\ref{rtoR}))
\begin{equation}\label{oprepl}
O({\bf r})\rightarrow O({\bf R})=\int d{\bf q}\, {\tilde O}(\bf
q)\exp{\left[i{\bf q}\cdot
   ({\bf r}+{\sqrt\theta}{\mbox{\boldmath $\rho$}})\right]}\, .
\end{equation} The properties of products is determined by studying
products of $\exp{i\left[{\bf q}\cdot({\bf
r}+{\sqrt\theta}{\mbox{\boldmath $\rho$}})\right]}$'s;
\begin{eqnarray}\label{Moyal4}
\exp\left[i{\bf k}\cdot({\bf r}
+{\sqrt\theta}{\mbox{\boldmath$\rho$}})\right]&\times& 
  \exp\left[i{\bf q}\cdot({\bf r}+{\sqrt\theta}{\mbox{\boldmath
  $\rho$}})\right]
\nonumber\\
&=&\exp\left[i({\bf k}+{\bf q})\cdot{\bf
r}\right]\exp(i\theta\epsilon_{ab}k_aq_b) 
  \exp\left[i{\sqrt\theta}({\bf k}+{\bf q})\cdot{\mbox{\boldmath
  $\rho$}}\right]\, .
\end{eqnarray} 
Taking the $H_{\rm aux}$ vacuum expectation value of
the above, equivalent to integrating out the $\rho$'s in
(\ref{genfunc4}), agrees with (\ref{Moyal3}).

It is straightforward to extend (\ref{Moyal4}) to more than two factors. 
\begin{equation}\label{multiprod}
\prod_n \exp\left[i{\bf k}_n\cdot({\bf r}
+{\sqrt\theta}{\mbox{\boldmath$\rho$}})\right]=\exp\left[ i\sum_n
{\bf k}_n\cdot{\bf r}\right]\exp\left[ i\theta\epsilon_{ab}
\sum_{n<m}k_{na}k_{mb}\right]
\exp\left[ i\sqrt{\theta}\sum_n{\bf
k}_n\cdot{\mbox{\boldmath$\rho$}}\right]\, .
\end{equation}
Although this product, before any expectation values are taken, is
associative, it ceases to be once the $H_{\rm aux}$ ground state
matrix elements are taken. Due to this non associativity, uniqueness
theorems \cite{unique} for deformations do not apply and this
deformation {\em is different} from the Groenewold-Moyal one.

The replacement (\ref{rtoR}) of ${\bf r}$ by ${\bf R}$ may appear
somewhat unnatural in that the $y$ coordinate is shifted by a a a
multiple of the coordinate $\rho_y$ while the coordinate $x$ is
shifted by a multiple of the conjugate momentum $\rho_x$. This may be
avoided by extending the auxiliary system to two dimensions,
considering {\mbox{\boldmath$\rho$}} as a coordinate vector,
introducing new conjugate momenta {\mbox{\boldmath$\pi$}}, with
$[\pi_a,\rho_b]=-i\delta_{ab}$, and placing the system in a strong
magnetic field. $H_{\rm aux}$ is replaced by
\begin{equation}\label{Hosc2}
H_{\rm aux}'=\frac{\pi_x^2}{2\mu}+\frac{(\pi_y-\rho_x)^2}{2\mu}+
  \frac{\omega}{2}\rho^2\, ;
\end{equation}
we have scaled the variables to set $eB=1$. The $\omega$ term breaks
the degeneracy of the lowest Landau level with energies in this
level being $E_n=\left[\sqrt{4\omega\mu+1}(n+1)-n\right]/(2\mu)$. 
The limit $\mu\rightarrow 0$, specifically $\mu<<1/\omega$, forces
this system into the lowest Landau level
\cite{Bigatti:1999iz,Jackiw:2001dj,Jackiw:2002wd}. $\rho_x$ does
become the momentum conjugate to $\rho_y$ and the preceding discussion
holds. For non zero $\mu$, $\exp i\omega|t-t'|$ in (\ref{Snc3}) is
replaced by $\left[\exp\left( i\omega|t-t'|\right)-
\exp \left(i|t-t'|/\mu\right)\right]$. At equal times $\rho_x$ and
$\rho_y$ commute and (\ref{rtoR}) doesn't introduce any equal time
coordinate noncommutativity. It is interesting to look at
$[x(t),x(t')]$ for $t\ne t'$. We have three regions of $|t-t'|$ to
consider. For $|t-t'|<\mu$ the commutator $[x(t),y(t')]\sim 0$; for
$\mu<|t-t'|<1/\omega$ the preceding discussions hold and the
commutator $\sim\theta$, while for $|t-t'|>1/\omega$ the dynamics of
$H_0$ determine the commutator.

Useful conversations with A.~Rajaraman are acknowledged.

\end{document}